\documentclass{LMCS}
\usepackage[T1]{fontenc}
\usepackage{microtype, enumerate, array, mathrsfs, amsthm, mathdefs, relsize, graphicx}
\makeatletter
\newcommand\m@thsm@ller[2]{\mbox{\relsize{-1}$\m@th#1#2$}}
\let\smaller\undefined
\DeclareRobustCommand\smaller[1]{\relax\ifmmode{\mathpalette\m@thsm@ller{#1}}\else{\relsize{-1}#1}\fi}
\makeatother

\newenvironment{enuma}{\begin{enumerate}[\upshape(a)]}{\end{enumerate}}
\newenvironment{enumi}{\begin{enumerate}[\upshape(i)]}{\end{enumerate}}
\newenvironment{enuM}{\begin{enumerate}[\hskip-4 pt]}{\end{enumerate}}

\newcommand*{\FO}{\smaller{\mathrm{FO}}}
\newcommand*{\MSO}{\smaller{\mathrm{MSO}}}
\newcommand*{\GSO}{\smaller{\mathrm{GSO}}}
\newcommand*{\WMSO}{\smaller{\mathrm{WMSO}}}
\newcommand*{\lex}{\mathrm{lex}}
\newcommand*{\?}{\kern .08em}

\newcommand\nsubseteq{\not\subseteq}
\newcommand\nsqsubseteq{\not\sqsubseteq}

\newcommand\upqed{\vskip-\baselineskip\vskip-\belowdisplayskip}

\def\doi{6 (1:4) 2010}
\lmcsheading%
{\doi}
{1--22}
{}
{}
{Jan.~11, 2008}
{Feb.~16, 2010}
{}   

\begin{document}
\title{Guarded Second-Order Logic, Spanning Trees, and Network Flows}
\author[A.\ Blumensath]{Achim Blumensath}
\address{TU Darmstadt, Germany}
\email{blumensath@mathematik.tu-darmstadt.de}


\begin{abstract}
According to a theorem of Courcelle monadic second-order logic
and guarded second-order logic (where one can also quantify over
sets of edges) have the same expressive power
over the class of all countable $k$-sparse hypergraphs.
In the first part of the present paper we extend this result
to hypergraphs of arbitrary cardinality.
In the second part,
we present a generalisation dealing with methods to encode
sets of vertices by single vertices.
\end{abstract}

\maketitle
\section*{introduction}

\noindent Guarded second-order logic ($\GSO$) is the variant of monadic second-order logic
($\MSO$) where one can not only quantify over sets of vertices but also over sets of edges.
This modification results in a large increase of expressive power.
Statements that can be expressed in guarded second-order logic, but not
in mo\-nad\-ic second-order logic, include the existence of certain minors in a graph
and the existence of Hamiltonian paths.

The high expressive power of guarded second-order logic means that most
$\GSO$-theories are quite complicated. In~\cite{Seese91} Seese has shown
that every class of graphs with infinite tree width has an undecidable $\GSO$-theory.
This result immediately generalises to hypergraphs.
It follows that all classes of hypergraphs with a decidable $\GSO$-theory
are \emph{$k$-sparse,} for some~$k$, which roughly means that their members have few edges.
For classes of countable $k$-sparse hypergraphs, Courcelle~\cite{Courcelle03} has shown
that every $\GSO$-formula is equivalent to an $\MSO$-formula over such a class.
It follows that over every class of countable hypergraphs with a decidable $\GSO$-theory
guarded second-order logic and monadic second-order logic have the same
expressive power.
Unfortunately, the proof of Theorem~1.4 in~\cite{Courcelle03} contains an error.
In the first part of the present article we give a new proof of this theorem.
In addition, we extend the result from countable hypergraphs
to hypergraphs of arbitrary cardinality.

When we look at the results of the first part we see that most of them
concern the coding of sets of vertices by single vertices.
In the abstract, this problem can be stated as follows\?:
given a set $F \subseteq \PSet(V)$ of finite sets of vertices,
find a definable function $f : F \to V$ that is injective.
In our concrete case, $F := E$ is the set of edges.
In the second part of the paper
we consider more general instances of this problem where $F$~can be arbitrary.
This generalisation is inspired by a result of
Colcombet and L\"oding~\cite{ColcombetLoeding07} on set interpretations.
Their main technical result is a method to transform
a definable finite-to-one function $F \to V$ into an injective one.
Colcombet and L\"odung consider as background structure only the infinite binary tree.
Below we show that using guarded second-order parameters,
i.e., sets of edges, we can extend some of their results to arbitrary graphs.

The overview of the article is a follows.
We start in Section~\ref{Sect: sparse orientations}
with basic definitions and a survey of results on definable
orientations of sparse hypergraphs.
In Section~\ref{Sect: spanning tree} we prove
the general version of the one technical result of~\cite{Courcelle03}
whose proof does not extend to arbitrary cardinalities.
In Section~\ref{Sect: GSO and MSO}
we summarise the consequences for the expressive power of guarded second-order
logic on sparse hypergraphs.

Section~\ref{Sect: flow} contains the second part of the article.
We study network flow problems and we show how to use flows
to transform definable finite-to-one maps into injective ones.

\section{Orientations of sparse hypergraphs}
\label{Sect: sparse orientations}

\noindent Let us fix our terminology regarding graphs and hypergraphs.
When we say `graph' we will mean an undirected one.
Undirected graphs will always be simple and loop free,
whereas directed graphs will be simple, but they may contain loops.
When dealing with hypergraphs we will sometimes allow multiple edges.
Such a hypergraph is a two-sorted structure $\langle V, E, I\rangle$
where $V$~is the set of vertices, $E$~the set of edges, and $I \subseteq V \times E$
the incidence relation. Using sloppy notation we will tacitly identify an edge $e \in E$
of such a hypergraph with the set $\set{ v \in V }{ \langle v,e\rangle \in I }$
of its vertices and we write $v \in e$ instead of $\langle v,e\rangle \in I$.
Similarly, if $F \subseteq E$ is a set of edges then the union~$\bigcup F$
consists of all vertices incident with at least one edge of~$F$. We will use
this notation even if there are multiple edges.

\emph{Monadic second-order logic} ($\MSO$) extends first-order logic by
variables and quantifiers that range over sets of vertices.
Similarly, \emph{guarded second-order logic} ($\GSO$) extends first-order logic
by variables and quantifiers ranging over sets of vertices or sets of edges
(for detailed definitions see \cite{GraedelHirschOtto02}).
We will also consider \emph{weak monadic second-order logic} ($\WMSO$)
where quantification is restricted to \emph{finite} sets of vertices.
\begin{defi}
Let $\frakH = \langle V,E\rangle$ be a hypergraph.
\begin{enumerate}[(a)]
\item We say that $\frakH$~has \emph{rank}~$m$ if every edge of~$\frakH$ has at most~$m$
vertices.

\item A \emph{subhypergraph} of~$\frakH$ is a hypergraph $\frakH_0 = \langle V_0,E_0\rangle$
with $V_0 \subseteq V$ and $E_0 \subseteq E$.

\item Let $C \subseteq V$.
The subhypergraph \emph{induced by~$C$} is
\begin{align*}
  \frakH|_C := \langle C, E|_C\rangle
  \qtextq{with}
  E|_C := \set{ e \in E }{ e \subseteq C }\,.
\end{align*}
\end{enumerate}
\end{defi}

In order to translate $\GSO$-formulae into $\MSO$-formulae, we have
to encode sets of edges by sets of vertices. A simple way to do
so consists in choosing an \emph{orientation} of the hypergraph, i.e.,
a function assigning to each edge one of its vertices.
\begin{defi}
Let $\frakH = \langle V,E\rangle$ be a hypergraph.
\begin{enumerate}[(a)]
\item
An \emph{orientation}%
\footnote{This is called a \emph{semi-orientation} in \cite{Courcelle03}.}
of~$\frakH$ is
a function $f : E \to V$ with $f(e) \in e$, for all $e \in E$.
We say that a formula~$\varphi(x,Y)$ \emph{defines} an orientation~$f$ of~$\frakH$
if we have
\begin{align*}
  \frakH \models \varphi(a,e) \quad\iff\quad f(e) = a\,,
  \quad\text{for all } a \in A \text{ and } e \in E\,.
\end{align*}

\item An orientation~$f$ is \emph{bounded by~$k$} if
\begin{align*}
  \abs{f^{-1}(a)} \leq k\,, \quad\text{for all } a \in A\,.
\end{align*}

\item We call~$\frakH$ \emph{$\MSO$-orientable} if there exist
an $\MSO$-formula~$\varphi(x,Y;\bar P)$ with parameters
$P_i \subseteq V$ defining an orientation of~$\frakH$.
Similarly, we say that $\frakH$~is \emph{$\GSO$-orientable} if there exist
a $\GSO$-formula~$\varphi(x,Y;\bar P,\bar Q)$ with parameters
$P_i \subseteq V$ and $Q_i \subseteq E$
defining an orientation of~$\frakH$.
\end{enumerate}
\end{defi}

\noindent In this paper we are mainly interested in \emph{sparse}
hypergraphs, i.e., hypergraphs with few edges.
\begin{defi}
A hypergraph $\frakH = \langle V,E\rangle$ is \emph{$k$-sparse}%
\footnote{In \cite{Courcelle03} such hypergraphs are called \emph{uniformly $k$-sparse.}
Courcelle also introduces a notion of a \emph{$k$-sparse} graph.
Since uniform sparsity is the more robust notion, and the only one we will use in this paper,
we have changed terminology for brevity. A related notion is the \emph{arboricity} of a graph
(see, e.g., Section 2.4 of~\cite{Diestel06}).}
if
\begin{align*}
  \bigabs{E|_X} \leq k\cdot \abs{X}\,,
  \quad\text{for every finite set } X \subseteq V\,.
\end{align*}
\end{defi}
\begin{lem}
Let $\frakG = \langle V,E\rangle$ be a graph.
\begin{enuma}
\item If the degree of\/~$\frakG$ is at most~$2k$, then\/ $\frakG$~is $k$-sparse.
\item If\/ $\frakG$~is planar, then it is $3$-sparse.
\end{enuma}
\end{lem}
\proof\hfill
\begin{enuma}
\item If $\frakX = \langle X,F\rangle$ is a finite induced subgraph of~$\frakG$ then
\begin{align*}
  2\cdot\abs{F} = \sum_{v \in X} \deg(v) \leq 2k\cdot\abs{X}\,.
\end{align*}

\item This follows from the fact that every planar graph with $n$~vertices
has at most $3n-6$ edges (see, e.g., Corollary~4.2.10 of~\cite{Diestel06}).
\qed
\end{enuma}

\noindent In the next section we will prove that every hypergraph of
bounded rank is $\GSO$-orientable.  In the remainder of this section
we show that $k$-sparse hypergraphs are even $\MSO$-orientable.  For
countable hypergraphs these results are all due to
Courcelle~\cite{Courcelle03}.  The only thing new in the present
section are two applications of the compactness theorem for
first-order logic to extend the results to uncountable hypergraphs.
The proofs in Section~\ref{Sect: spanning tree}, on the other hand,
are mostly new.
\begin{lem}\label{Lem: k-sparse hypergraphs and orientations}
A hypergraph $\frakH = \langle V,E\rangle$ \textup(possibly with multiple edges\textup)
of finite rank is $k$-sparse if and only if
there exists an orientation of~$\frakH$ that is bounded by~$k$.
\end{lem}
\proof
For $(\Leftarrow)$, let $X \subseteq V$ be finite.
Then
\begin{align*}
  \bigabs{E|_X} \leq \sum_{a \in X} \abs{f^{-1}(a)} \leq k\cdot\abs{X}\,.
\end{align*}
$(\Rightarrow)$ First, let us consider the case where $\frakH$~is finite.
If $f$~is an arbitrary orientation of~$\frakH$ then
\begin{align*}
  \sum_{a \in V} \abs{f^{-1}(a)} = \abs{E} \leq k\cdot\abs{V}\,.
\end{align*}
Hence, if there is some element $a \in V$ with $\abs{f^{-1}(a)} > k$
then there must be some other element $b \in V$ with
$\abs{f^{-1}(b)} < k$.
Let us define the \emph{weight} of an orientation by
\begin{align*}
  w(f) := \sum {\bigset{ \abs{f^{-1}(a)} - k }{ a \in V,\ \abs{f^{-1}(a)} > k }}\,.
\end{align*}
We have to construct an orientation of weight~$0$.
To do so we transform an orientation~$f$ with $w(f) > 0$ into
one with smaller weight.
Given~$f$, fix an element $a \in V$ with $\abs{f^{-1}(a)} > k$.
Let $F \subseteq E$ be the smallest subset of~$E$ such
that $a$~belongs to the set $U := \bigcup F$ and
we have $f^{-1}(c) \subseteq F$, for every element $c \in U$.
The subhypergraph~$\frakH|_U$ induced by~$U$ is $k$-sparse. Hence,
there exists some element $b \in U$ with $\abs{f^{-1}(b)} < k$.
By choice of~$F$ we can find a sequence of edges
$e_0,\dots,e_n \in F$ with
\begin{align*}
  b \in e_0\,,\quad
  f(e_i) \in e_{i+1}\,,
  \quad\text{and}\quad
  f(e_n) = a\,.
\end{align*}
We define a new orientation~$g$ by setting
\begin{align*}
  g(e) :=
  \begin{cases}
    b          &\text{if } e = e_0\,, \\
    f(e_{i-1}) &\text{if } e = e_i,\ i > 0\,, \\
    f(e)       &\text{otherwise}\,.
  \end{cases}
\end{align*}
It follows that
\begin{align*}
   \abs{g^{-1}(x)} =
   \begin{cases}
     \abs{f^{-1}(a)} - 1 &\text{if } x = a\,,\\
     \abs{f^{-1}(b)} + 1 &\text{if } x = b\,,\\
     \abs{f^{-1}(x)}     &\text{otherwise}\,.
   \end{cases}
\end{align*}
Hence, $w(g) < w(f)$. Repeating this construction we obtain an orientation~$f$
with $w(f) = 0$.

The general case where $\frakH$~may be infinite can be proved using
the compactness theorem for first-order logic.
Let $\Delta$~be the elementary diagram of~$\frakH$ (i.e., the set of all
first-order formulae with parameters that hold in~$\frakH$\?; see \cite{Hodges93} for details)
where we consider~$\frakH$
as a two-sorted structure $\langle V, E, I\rangle$ with a binary incidence relation~$I$.
We can write down a formula~$\varphi$ stating that
$f : E \to V$ is a function such that
\begin{enumerate}[$\bullet$]
\item $\langle f(e), e\rangle \in I$\,, for all $e \in E\,,$
\item $\abs{f^{-1}(a)} \leq k$\,, for all $a \in V\,.$
\end{enumerate}
By assumption and the first part of the proof,
every finite subset of $\Delta \cup \{\varphi\}$ is satisfiable.
Therefore, according to the compactness theorem, there exists a model
$\frakH^+\!=\!\langle V^+\!,E^+\!,I^+\!,f^+\rangle$ of $\Delta \cup \{\varphi\}$.
By the Diagram Lemma (see, e.g., \cite{Hodges93}),
we can find an elementary embedding $h : \frakH \to \frakH^+$
(i.e., an embedding preserving every first-order formula).
Since every edge of~$\frakH$ has only finitely many vertices
it follows that
\begin{align*}
  \langle a, h(e)\rangle \in I^+
  \qtextq{implies}
  a = h(v)\,,\quad\text{for some } v \in e\,.
\end{align*}
Hence, we can define the desired orientation of~$\frakH$ by
$f := h^{-1} \circ f^+ \circ h$.
\qed

It turns out that the orientation obtained via the preceding lemma is
$\MSO$-definable. The following sequence of lemmas shows how we can encode
such an orientation by a finite set of unary predicates.
\begin{defi}
Let $\frakH = \langle V,E\rangle$ be a directed graph and $\frakG$
an undirected one.
\begin{enuma}
\item Every orientation~$f$ of~$\frakG$ induces an directed graph~$\frakG_f$
by orienting every edge~$e$ of~$\frakG$
such that it points to the vertex~$f(e)$.

\item An \emph{$\frakH$-orientation} of~$\frakG$
consists of a pair $\langle f,h\rangle$
where $f$~is an orientation of~$\frakG$
and $h$~is a homomorphism $\frakG_f \to \frakH$.

We say that an $\frakH$-orientation $\langle f,h\rangle$
is \emph{bounded by~$k$} if $f$~is bounded by~$k$.

\item We say that a family~$(P_v)_{v \in V}$ of unary predicates
\emph{encodes} an $\frakH$-orien\-ta\-tion $\langle f,h\rangle$
of~$\frakG$ if $P_v = h^{-1}(v)$, for all $v \in V$.
\end{enuma}
\end{defi}

\begin{lem}\label{Lem: encoding of H-orientations definable}
For every finite graph~$\frakH$,
there exists a first-order formula~$\varphi_\frakH(\bar X)$ such that
\begin{align*}
  \frakG \models \varphi_\frakH(\bar P)
  \quad\iff\quad
  \text{the tuple } \bar P \text{ encodes an $\frakH$-orientation of\/ } \frakG\,.
\end{align*}
\end{lem}
\proof
Let $u_0,\dots,u_{n-1}$ be an enumeration of the vertices of~$\frakH$.
All $\varphi_\frakH(\bar X)$ has to say is that
the~$X_i$ form a partition of the vertices (some~$X_i$ may be empty)
and that there is no edge $\{v,w\}$ of~$\frakG$
such that $v \in X_i$, $w \in X_k$ and $\langle u_i,u_k\rangle$
is \emph{not} an edge of~$\frakH$.
\qed

\begin{thm}[Ne\v set\v ril, Sopena, Vignal \cite{NesetrilSopenaVignal97}]
For every $k < \omega$, there exists a finite loop-free directed graph~$\frakT_k$ with
antisymmetric edge relation that has the following property.
For every finite directed graph\/~$\frakG$,
with irreflexive and antisymetric edge relation and indegree at most~$k$,
there exists a homomorphism $\frakG \to \frakT_k$.
\end{thm}
\begin{cor}\label{Cor: existence of bounded orientations}
Every $k$-sparse undirected graph has a $\frakT_k$-orientation which is bounded by~$k$.
\end{cor}
\proof
In Lemma~\ref{Lem: k-sparse hypergraphs and orientations},
we have shown that such a graph $\frakG = \langle V,E\rangle$ has an orientation $f : E \to V$
that is bounded by~$k$.
It follows that $\frakG_f$~has indegree at most~$k$.
By the theorem, there exists a homomorphism $h : \frakG_f \to \frakT_k$.
Thus, $\langle f,h\rangle$ is the desired $\frakT_k$-orientation.
\qed

\begin{lem}\label{Lem: T-orientations definable}
For every $k < \omega$, there exists a first-order formula~$\eta_k(\bar X)$
such that
\begin{align*}
  \frakG \models \eta_k(\bar P)
  \quad\iff\quad
  \bar P \text{ encodes a $\frakT_k$-orientation of\/ } \frakG \text{ that is bounded by } k\,.\!\!\!
\end{align*}
\end{lem}
\proof
Note that the homomorphism~$h$ of a $\frakT_k$-orientation $\langle f,h\rangle$
uniquely determines the orientation~$f$ since the edge relation of~$\frakT_k$ is antisymmetric.
In particular, the parameters~$\bar P$ encoding $\langle f,h\rangle$
tell us whether $f$~is bounded by~$k$.
Hence, we can obtain~$\eta_k(\bar X)$ by adding
a check for boundedness to the formula~$\varphi_{\frakT_k}(\bar X)$
of Lemma~\ref{Lem: encoding of H-orientations definable}.
\qed

\begin{cor}
The class of all $k$-sparse undirected graphs is finitely $\MSO$-axioma\-ti\-s\-able.
\end{cor}
\proof
By Lemma~\ref{Lem: k-sparse hypergraphs and orientations} and
Corollary~\ref{Cor: existence of bounded orientations}
it follows that a graph~$\frakG$ is $k$-sparse if and only if
it has a $\frakT_k$-orientation that is bounded by~$k$.
Hence, we can use the formula $\exists\bar X\eta_k(\bar X)$ where
$\eta_k$~is the formula from Lemma~\ref{Lem: T-orientations definable}.
\qed

In order to apply these results to hypergraphs we use the following
construction associating a graph with every hypergraph.
\begin{defi}
Let $\frakH = \langle V,E\rangle$ be a hypergraph with orientation~$f$.
We define a directed graph $\calO_f(\frakH) := \langle V,F\rangle$
with edge relation
\begin{align*}
  F := \set{ \langle a,b\rangle }
           { a \neq b \text{ and there is some edge } e \in E \text{ with }
             a \in e \text{ and } f(e) = b }\,.
\end{align*}
\end{defi}

\begin{lem}
Let\/ $\frakH = \langle V,E\rangle$ be a $k$-sparse hypergraph of rank~$m$
where $0 < k < \omega$ and $1 < m < \omega$.
Then\/ $\frakH$~has an orientation~$f$ that is bounded by~$mk^2$ such that
the edge relation of $\calO_f(\frakH)$ is antisymmetric.
\end{lem}
\proof
First, we consider the case that $\frakH$~is finite.
We call an element $a \in V$ \emph{bad} for an orientation~$f$ of~$\frakH$
if there is some element~$b \in V$
such that $\calO_f(\frakH)$ contains both edges $\langle a,b\rangle$ and $\langle b,a\rangle$.
Note that this implies that the vertex~$b$ is also bad.

We construct a sequence of orientations~$(f_n)_n$ such that
\begin{align*}
  \abs{f_n^{-1}(a)} \leq
  \begin{cases}
    k    &\text{if $a$ is bad for } f_n\,, \\
    mk^2 &\text{otherwise}\,,
  \end{cases}
\end{align*}
and the number of bad elements decreases at every step.
We start with an arbitrary orientation~$f_0$ bounded by~$k$.

Given an orientation~$f_n$ with the above properties we construct
a new orientation~$f_{n+1}$ with fewer bad elements as follows.
Let $a$~be a bad element, set $X := f_n^{-1}(a)$, and let
\begin{align*}
  Y := \bigset{ e }{ \textstyle a \in e \text{ and } f_n(e) \in \bigcup X \setminus \{a\} }\,.
\end{align*}
Since $a$~is bad we have
\begin{align*}
  \abs{X} \leq k
  \quad\text{and}\quad
  \abs{\textstyle\bigcup X} \leq k(m-1)\,.
\end{align*}
Note that every element of the form $b := f_n(e)$ with $e \in Y$ is also bad since,
by definition of~$X$, there is an edge $e' \in X$ with
\begin{align*}
  b \in e'
  \quad\text{and}\quad
  f_n(e') = a\,.
\end{align*}
Consequently, $\calO_{f_n}(\frakH)$ contains the edges $\langle b,a\rangle$ (since $f_n(e') = a$)
and $\langle a,b\rangle$ (since $f_n(e) = b$).
It follows that
\begin{align*}
  \abs{Y} \leq k\cdot\abs{\textstyle\bigcup X \setminus \{a\}} \leq k^2(m-1)\,.
\end{align*}
We define the new orientation~$f_{n+1}$ by
\begin{align*}
  f_{n+1}(e) := \begin{cases}
                  a      &\text{if } e \in Y\,, \\
                  f_n(e) &\text{otherwise}\,.
                \end{cases}
\end{align*}
Then we have
\begin{align*}
  \abs{f_{n+1}(x)^{-1}} \leq \begin{cases}
                               k + k^2(m-1)      &\text{if } x = a\,, \\
                               \abs{f_n(x)^{-1}} &\text{otherwise}\,.
                             \end{cases}
\end{align*}
In particular, $f_{n+1}$~is bounded by~$mk^2$.
By construction, the element~$a$ is not bad for~$f_{n+1}$.
Furthermore, if $\langle b,c\rangle$ is an edge in $\calO_{f_{n+1}}(\frakH)$
with $b,c \neq a$ then this edge is induced by an edge~$e$ in~$\frakH$
with $e \notin X \cup Y$.
Hence, $\langle b,c\rangle$ is also an edge of~$\calO_{f_n}(\frakH)$.
Therefore, every element that is bad for $f_{n+1}$ is also bad for~$f_n$.

It remains to prove the claim for infinite hypergraphs~$\frakH$.
Let $\Phi$~be the union of the elementary diagram of~$\frakH$
and formulae stating that $f$~is an orientation of~$\frakH$
that is bounded by~$mk^2$ and that $\calO_f(\frakH)$ has an antisymmetric
edge relation.
If $\frakM$~is a model of~$\Phi$ then there exists an embedding
$h : \frakH \to \frakM$ and the desired orientation of~$\frakH$
can be obtained via~$h$ from that of~$\frakM$.
Hence, it is sufficient to show that $\Phi$~is satisfiable.
Note that every finite subset $\Phi_0 \subseteq \Phi$ is satisfiable
since every finite substructure of~$\frakH$ has an orientation
of the desired form. By the compactness theorem it follows that
$\Phi$~is satisfiable.
\qed

\section{Depth-first spanning trees}   
\label{Sect: spanning tree}

\noindent While $k$-sparse hypergraphs are $\MSO$-orientable
there are hypergraphs without an $\MSO$-definable orientation.
For instance, the countably infinite clique is such a graph.
In this section we will show that every hypergraph of bounded rank
is at least $\GSO$-orientable.
A basic tool the proof below is based on is the notion of a spanning tree
of a hypergraph.
Before presenting the rather involved definition for hypergraphs
let us start with considering the simpler case of graphs.

For a countable undirected graph~$\frakG$ we can define a \emph{depth-first spanning tree}
to be a spanning tree~$\frakT$ of~$\frakG$ where no edge of~$\frakG$ connects
disjoint subtrees of~$\frakT$ (see \cite{Courcelle03,Diestel06}\?;
in~\cite{Diestel06} such trees are called \emph{normal}).
To generalise this definition to uncountable graphs we have to admit
trees of arbitrary ordinal height.
Such trees are necessarily \emph{order trees,} i.e., partial orders
$\langle T, {\leq}\rangle$ where $\leq$~is a \emph{tree order,}
that is, a partial order such that any two elements have an infimum and,
for every element~$a$, the set of all elements below~$a$ is well-ordered.
Unfortunately, we cannot in general hope to have a spanning subgraph
that is an order tree, since the partial order~$\leq$ requires too many edges.
Therefore, we will use a hybrid between an ordinary tree and an order tree.
The precise definition of a spanning tree~$\frakT$ of a graph~$\frakG$ is as follows.
Instead of requiring~$\frakT$ to be a subgraph
of~$\frakG$ we consider trees~$\frakT$ such that
\begin{enumerate}[$\bullet$]
\item for every vertex~$w$ of~$\frakT$ with immediate predecessor~$v$, the
  edge $\langle v,w\rangle$ belongs to~$\frakG$, and
\item for every vertex~$w$ of~$\frakT$ without immediate predecessor, we can fix
  an increasing chain $(u_i)_{i < \alpha}$ of predecessors of~$w$
  with limit~$w$ and a family $(\pi_i)_{i < \alpha}$ of paths from~$w$ to~$u_i$.
\end{enumerate}
Hence, every vertex~$w$ of~$\frakT$ is attached to its predecessors via
some auxiliary graph~$F_w$ that is either a single edge or a tree with root~$w$
whose leaves form an increasing sequence of predecessors of~$w$ with limit~$w$.
\begin{exa}
Consider the complete graph $\frakK_\kappa$, for an uncountable cardinal~$\kappa$.
We can enumerate the vertices of~$\frakK_\kappa$ as $(v_\alpha)_{\alpha < \kappa}$
where the index~$\alpha$ ranges over all ordinals less than~$\kappa$.
As depth-first spanning tree of this graph
we can use a chain of length~$\kappa$ as follows.
We set $\frakT := \langle T,E\rangle$ where
\begin{align*}
  T := \set{ v_\alpha }{ \alpha < \kappa }
\end{align*}
is the set of all vertices and
\begin{align*}
  E := \set{ \langle v_\alpha,v_{\alpha+1}\rangle }{ \alpha < \kappa }
 \cup \set{ \langle v_\alpha,v_\delta\rangle }{ \delta \text{ a limit ordinal and } \alpha < \delta }\,.
\end{align*}
The first part of~$E$ consists of the successor edges, whereas the second part
contains the auxiliary graphs~$F_{v_\delta}$ attaching a limit vertex~$v_\delta$ to its predecessors.
\end{exa}

To generalise these ideas to hypergraphs we need a suitable replacement
for the trees~$F_w$. Unfortunately, not every hypergraph has a spanning tree.
A typical example is the hypergraph
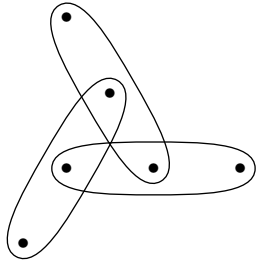
\begin{center}
\includegraphics{Sparse-final.1}
\end{center}
Instead, we will use certain tree-like hypergraphs called \emph{priority trees.}
\begin{defi}
Let $\frakH = \langle V,E\rangle$ be a hypergraph.
A \emph{hyperpath} in~$\frakH$ is a sequence $e_0\dots e_n$ of edges such that
\begin{align*}
  e_i \cap e_k \neq \emptyset \quad\iff\quad \abs{i - k} \leq 1\,.
\end{align*}
If $u \in e_0 \setminus e_1$ and $v \in e_n \setminus e_{n-1}$ then we say that
the hyperpath \emph{connects $u$~and~$v$.}
\end{defi}
\begin{defi}
Let $\frakH = \langle V,E\rangle$ be a hypergraph of rank at most~$m$,
$\langle T, F\rangle$ be a subhypergraph of~$\frakH$
with $T = \bigcup F$,
and suppose that there are partitions
\begin{align*}
  T = P_0 \cupdot\dots\cupdot P_{m-1}
  \qtextq{and}
  F = F_0 \cupdot\dots\cupdot F_{m-1}\,.
\end{align*}
\begin{enuM}
\item(a)\enspace
Suppose that $\frakT = \langle T, F, L, (F_i)_{i < m}, (P_i)_{i < m}, v\rangle$
with $L \subseteq F$ and $v \in T$.
We define by induction when such a tuple~$\frakT$ is a \emph{priority tree.}
The element~$v$ is called the \emph{root} of~$\frakT$ and $L$~is its set of \emph{leaf edges.}

We start the induction with the case where
$F$~consists of a single hyperpath $e_0\dots e_n$ with $v \in e_0 \setminus e_1$,
we have $L = \{e_n\}$, $F_0 = F$, and $P_0 = T$.
Then $\frakT$~is a priority tree.
We also call $\frakT$~a priority tree if it
can be obtained from a priority tree $\frakT' = \langle T',F',L',(F'_i)_i,(P'_i)_i,v\rangle$
with the same root~$v$ by adding a hyperpath $e_0\dots e_n$ such that
\begin{align*}
  &e_i \cap T' \neq \emptyset \quad\iff\quad i = 0\,, \\
  &e_0 \nsubseteq T'\,, \\
  &L = L' \cup \{e_n\}\,, \\
  &e_0,\dots,e_n \in F_k\,, \\
  &(e_0 \cup\dots\cup e_n) \setminus T' \subseteq P_k\,,
\end{align*}
where $k$~is the minimal index such that $e_0 \cap P'_k = \emptyset$.
This is the successor case of the induction step.

Finally, we also have a limit case. Suppose that
\begin{align*}
  \frakT^\alpha = \langle T^\alpha, F^\alpha, L^\alpha, (F^\alpha_i)_i, (P^\alpha_i)_i,v\rangle\,,
  \qquad\text{for } \alpha < \beta\,,
\end{align*}
is an increasing chain of priority trees. That is,
the sequences $(T^\alpha)_\alpha$, $(F^\alpha)_\alpha$, $(L^\alpha)_\alpha$,
$(F^\alpha_i)_\alpha$, and $(P^\alpha_i)_\alpha$ are all increasing,
and all trees~$\frakT^\alpha$ have the same root~$v$.
Then $\frakT$~is a priority tree if it is the union of this chain,
that is, if
\begin{align*}
  T = \bigcup_{\alpha < \beta} T^\alpha,\quad
  F = \bigcup_{\alpha < \beta} F^\alpha,\quad
  L = \bigcup_{\alpha < \beta} L^\alpha,\quad
  F_i = \bigcup_{\alpha < \beta} F_i^\alpha,\quad
  P_i = \bigcup_{\alpha < \beta} P_i^\alpha.
\end{align*}

\item(b)\enspace A \emph{branch} of~$\frakT$ is a hyperpath $e_0\dots e_m \subseteq F$
satisfying the following conditions\?:
\begin{enumerate}[$\bullet$]
\item $e_0 \setminus e_1$ contains the root~$v$ of~$T$.
\item Let $k_i$~be the index such that $e_i \in F_{k_i}$.
  We have $e_{i+1} \setminus e_i \subseteq P_{k_{i+1}}$, for every $i < m$.
  Furthermore, if $k_i \neq k_{i+1}$ then $k_i := \min {\set{ l }{ e_{i+1} \cap P_l \neq \emptyset }}\,.$
\end{enumerate}

\item(c)\enspace With each priority tree~$\frakT$ we associate two relations,
an order~$\leq$ on~$F$ defined by
\begin{align*}
  e \leq f \quad\defiff\quad \text{every branch containing~$f$ also contains~$e$,}
\end{align*}
and an equivalence relation~$\sim$ on~$T$ defined by
\begin{align*}
  u \sim v \quad\defiff\quad
  & u,v \in P_k\,, \text{ for some } k\,, \text{ and there exists a hyperpath} \\
  & e_0\dots e_m \subseteq F_k \text{ connecting } u \text{ and } v\,.
\end{align*}
\end{enuM}
\end{defi}
\begin{exa}
Consider the following priority tree with edges $a,b,c,d,e,f$
where we have labelled each vertex in~$P_i$ by the index~$i$.
The edge colours are given by $F_0 = \{a,b,e\}$, $F_1 = \{c\}$, $F_2 = \{d,f\}$.
The ordering~$\leq$ is displayed to the right.
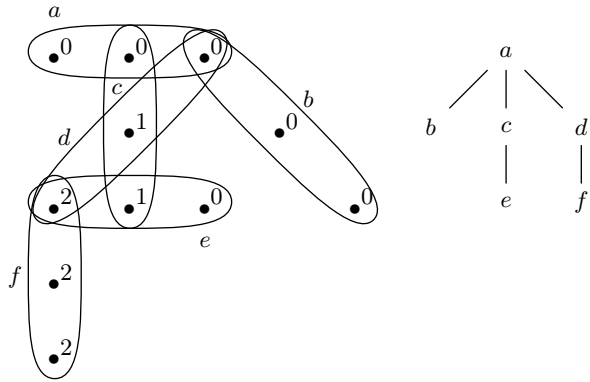
\begin{center}
\includegraphics{Sparse-final.2}
\end{center}
\end{exa}

Recall that a \emph{tree order} is a partial order such that
any two elements have an infimum and,
for every element~$a$, the set of all elements below~$a$ is well-ordered.
A~\emph{preorder} is a reflexive and transitive relation.
Every preorder~$\sqsubseteq$ induces an equivalence relation
${\sqsubseteq} \cap {\sqsubseteq^{-1}}$.
The equivalence classes of this relation are called \emph{$\sqsubseteq$-classes.}
\begin{lem}
Let $\frakT$~be a priority tree. The order~$\leq$ on the edges is a tree order.
\end{lem}
The proof consists of a straightforward but tedious induction following the
construction of priority trees.

\begin{lem}\label{Lem: existence of priority trees}
Let $\frakH = \langle V,E\rangle$ be a connected hypergraph of rank at most~$m$.
\begin{enuma}
\item
For each vertex $v \in V$ and every set $L_0 \subseteq E$ of edges,
there exists a priority tree $\frakT = \langle T, F, L, (F_i)_i, (P_i)_i, v\rangle$
with root~$v$ such that $\bigcup L_0 \subseteq T$ and $L \subseteq L_0$.

\item For every $\GSO$-formula $\vartheta(x,y)$ \textup(possibly with parameters\textup),
there exists a $\GSO$-formula $\varphi(x,y)$ \textup(with parameters\textup) such that,
if $\vartheta$~defines a well-order on~$L_0$ and $\frakT$~is a priority tree as in~\itm{(a)}
then $\varphi(x,y)$~defines a linear order on~$T$.
\end{enuma}
\end{lem}
\proof\hfill
\begin{enuM}
\item(a)\enspace Let $(e_i)_{i < \alpha}$ be an enumeration of~$L_0$.
For every $i < \alpha$, we fix a hyperpath $\pi_i = h^i_0\dots h^i_{m_i}$
connecting~$v$ with $h^i_{m_i} = e_i$.
We construct~$\frakT$ by induction on~$i$. We start with the hyperpath~$\pi_0$.
At step $i > 0$ we determine the shortest suffix $h^i_l\dots h^i_{m_i}$ of the path~$\pi_i$ that
meets the tree constructed so far and we add this suffix to the tree.
(If $e_i \subseteq T$ we leave the tree unchanged.)
We choose the least index~$k$ with $h^i_l \cap P_k = \emptyset$ and we put
the new edges into~$F_k$ and the new vertices into~$P_k$.
The limit of this construction is the desired priority tree.

\item(b)\enspace The equivalence relation~$\sim$ associated with~$\frakT$ is $\MSO$-definable
in~$\frakH$ with the help of the parameters $T$,~$F$, $F_i$, and~$P_i$.
We denote the $\sim$-class of a vertex~$u$ by~$[u]$.
Note that, by construction of~$\frakT$,
$[u]$~is a hyperpath and $[u]$~contains a unique leaf edge which, furthermore,
is one of the ends of the hyperpath. We denote by~$\eta(u)$ the suffix of the hyperpath~$[u]$
that connects~$u$ to the leaf edge in~$[u]$.

To define the desired order on~$T$ we first
construct a preorder on~$T$ by setting $x \sqsubseteq y$ if and only if one of the following conditions is
satisfied\?:
\begin{enumerate}[$\bullet$]
\item $x \in P_i$ and $y \in P_k$, for $i < k$.
\item $x,y \in P_k$ and the leaf edge in~$[x]$ is $\vartheta$-smaller
  than the leaf edge in~$[y]$.
\item $x,y \in P_k$, $[x] = [y]$, and $\eta(x) \subseteq \eta(y)$.
\end{enumerate}
Note that we have $x \sqsubseteq y$ and $y \sqsubseteq x$ if and only if
$\eta(x) = \eta(y)$. In this case $x$~and~$y$ belong to the same edge $e \in F$.
Hence, every $\sqsubseteq$-class has size at most~$m$.
Adding $m$~additional unary predicates $Q_0,\dots,Q_{m-1}$ such that each~$Q_i$
contains at most one element of each $\sqsubseteq$-class, we can define
\begin{align*}
  x < y \quad\defiff\quad
    x \sqsubseteq y \text{ and }
    (&\text{either } y \nsqsubseteq x\,, \text{ or we have} \\
     &x \in Q_i \text{ and } y \in Q_k\,, \text{ for } i < k)\,.
\end{align*}
\upqed
\qed
\end{enuM}

\noindent We have seen that every $k$-sparse graph has an
$\MSO$-definable orientation~$f$ that is bounded by~$k$.  If we want
to encode sets of edges via sets of vertices we can try to encode each
edge~$e$ by a pair $\langle v,i\rangle$ consisting of the vertex $v :=
f(e)$ and a number $i < k$.  This idea requires a way to linearly
order the sets $f^{-1}(v)$.  In \cite{Courcelle03} Courcelle uses
depth-first spanning trees to obtain such linear orders.  As remarked
above one needs to adapt the definition of a depth-first spanning tree
when one tries to extend these results to uncountable hypergraphs.
\begin{defi}
Let $\frakH = \langle V,E\rangle$ be a hypergraph of rank~$m$
and suppose that $\frakT = \langle T, {\leq}, (F_v)_{v \in T}\rangle$
is a structure where
$\langle T, {\leq}\rangle$ is a tree (of ordinal height) with $T \subseteq V$ and
with every vertex $v \in T$ we associate a set $F_v \subseteq E$ of edges.
We assume that $F_u \cap F_v = \emptyset$, for $u \neq v$.
\begin{enuM}
\item(a)\enspace The set of \emph{auxiliary nodes} associated to a vertex $v \in T$ is
\begin{align*}
  A_v := \{v\} \cup \bigcup F_v \setminus \bigcup_{x < v} A_x\,.
\end{align*}

\item(b)\enspace For $X \subseteq V$, we define
\begin{align*}
  B(X/T) := \set{ v \in T }{ X \cap A_v \neq \emptyset }
\qtextq{and}
  \beta(X/T) := \max B(X/T)\,.
\end{align*}

\item(c)\enspace $\frakT$~is a \emph{depth-first spanning tree} of~$\frakH$
if it satisfies the following conditions\?:
\begin{enumerate}[$\bullet$]
\item For all $u \neq v$, $A_u \cap A_v = \emptyset$ and $A_u \cap T = \{u\}$.
\item For each edge $e \in E$ the set $B(e/T)$
  is nonempty and linearly ordered by~$\leq$.
\item The vertices $v \in T$ are partitioned into the following classes\?:
  $(0)$ the root\?; $(s_l)_{l < m}$ a successor\?; $(t_l)_{l < m}$ a limit\?;
  where the successor and limit vertices are subdivided into $m$ subclasses.
  This partition satisfies the following conditions\?:
  \begin{enumerate}[$s_i$]
  \item[$(0)$] If $v$~has type~$0$ then it is the root of~$T$ and $F_v = \emptyset$.
  \item[$(s_l)$]
    If $v$~has type~$s_l$ then it is the (immediate) successor of some vertex $u \in T$.
    We have $F_v = \{e\}$ with $v \in e$.
    Futhermore, $v$~is the only vertex in $B(e/T)$ of type~$s_l$
    and $\beta(e \setminus \{v\}/T) = u$.
  \item[$(t_l)$] If $v$~has type~$t_l$ then it is the limit
    of an increasing sequence $(u_i)_{i < \gamma}$ of vertices $u_i \in T$.
    $F_v$~is (the set of edges of) a priority tree with root~$v$.
    Furthermore,
    \begin{align*}
      \set{ \beta(e/T) }{ e \text{ a leaf edge of } F_v }
    \end{align*}
    is a cofinal subset of $(u_i)_i$ and $v$~is the only vertex in $B(\bigcup F_v/T)$
    with type~$t_l$.
  \end{enumerate}
\end{enumerate}
\end{enuM}
\end{defi}

\begin{prop}
Every connected hypergraph~$\frakH$ has a depth-first spanning tree.
\end{prop}
\proof
If in the definition of a depth-first spanning tree we drop the condition
that $B(e/T) \neq \emptyset$, for every edge~$e$, then we obtain
a structure that we call a \emph{partial} depth-first spanning tree.
We construct an increasing sequence
\begin{align*}
  \langle T_\alpha, {\leq}, (F_v)_{v \in T_\alpha}\rangle\,,
  \quad \alpha < \kappa\,,
\end{align*}
of such partial depth-first spanning trees with the property that, for every connected
component~$C$ of $U_\alpha := V \setminus \bigcup_{v \in T_\alpha} A_v$,
the set
\begin{align*}
  N(C/T_\alpha) := \bigcup {\set{ B(e/T_\alpha) }{ e \in E \text{ with } e \cap C \neq \emptyset }}
\end{align*}
is linearly ordered by~$\leq$.
(A \emph{connected component} of~$U_\alpha$ is a maximal subset $C \subseteq U_\alpha$ such that
the subhypergraph~$\frakH|_C$ is connected.)
The limit of this sequence will be the desired depth-first
spanning tree of~$\frakH$.

We start by choosing an arbitrary element $v \in V$ and setting
$T_0 := \{v\}$ and $F_v := \emptyset$.
For limit ordinals~$\delta$, we define $T_\delta := \bigcup_{\alpha < \delta} T_\alpha$.
For the successor step, suppose that we have already defined~$T_\alpha$.
Fix some connected component~$C$ of~$U_\alpha$.
Note that $N(C/T_\alpha)$ is nonempty since $\frakH$~is connected.
We distinguish two cases.
\begin{enumerate}[(1)]
\item If $N(C/T_\alpha)$ has a maximal element~$u$ then we choose some edge~$e$
with $e \cap A_u \neq \emptyset$ and $e \cap C \neq \emptyset$,
and we fix some vertex $v \in e \cap C$.
We add~$v$ to~$T_\alpha$ as immediate successor of~$u$ and
we set $F_v := \{e\}$.
It follows that $A_v = e \cap U_\alpha$.
Since $B(e/T_\alpha)$ contains at most $\abs{e \setminus \{v\}} < m$ vertices
there is some $l < m$ such that $B(e/T_\alpha)$ contains no vertex of type~$s_l$.
Hence, in the new tree $T_{\alpha+1}$ we can assign the type~$s_l$ to~$v$.

\item Suppose that $N(C/T_\alpha)$ has no maximal element. We choose a sequence
$(e_i)_{i < \gamma}$ of edges with
$e_i \cap C \neq \emptyset$ such that
the sequence $(u_i)_{i < \gamma}$ defined by
\begin{align*}
  u_i := \beta (e_i/T_\alpha)
\end{align*}
is increasing and cofinal in $N(C/T_\alpha)$.
By taking a suitable subsequence we may assume that
the set of types appearing in $B(e_i/T_\alpha)$ is the same for every $i < \gamma$.

For each edge~$e_i$, choose some edge $h_i \subseteq C$ with $h_i \cap e_i \neq \emptyset$
and set $L := \set{ h_i }{ i < \gamma }$.
We select a vertex $v \in C$ and a priority tree
$\frakS = \langle S, H, L, \bar F,\bar P,v\rangle$ such that $S \subseteq C$.
We define $T_{\alpha+1} := T_\alpha \cup \{v\}$ where $v$~is the limit of $N(C/T_\alpha)$ and
we set $F_v := H \cup \set{ e_i }{ i < \gamma }$.
It follows that $A_v = S \cup \bigcup_i (e_i \cap U_\alpha)$.
\end{enumerate}
It remains to show that the constructed tree $T_{\alpha+1}$ is a partial depth-first tree
where all sets $N(C/T_{\alpha+1})$ are linearly ordered.
We start by showing that each set $B(e/T_{\alpha+1})$ with $e \in E$ is linearly ordered.
If $e \cap A_v = \emptyset$ then $B(e/T_{\alpha+1}) = B(e/T_\alpha)$ and we are done.
Otherwise, we have $B(e/T_{\alpha+1}) = B(e/T_{\alpha+1}) \cup \{v\}$.
Note that $A_v \subseteq C$ implies $e \cap C \neq \emptyset$. Therefore, we have
$B(e/T_\alpha) \subseteq N(C/T_\alpha)$. Since $v$~is larger than every element in $N(C/T_\alpha)$
the claim follows.

Let $D$~be a connected component of $U_{\alpha+1} := V \setminus \bigcup_{x \in T_{\alpha+1}} A_x$.
We have to show that $N(D/T_{\alpha+1})$ is linearly ordered.
Since $U_{\alpha+1} \subseteq U_\alpha$ there is some connected component~$D'$ of~$U_\alpha$ containing~$D$.
If $D' \neq C$ then $U_\alpha \setminus U_{\alpha+1} \subseteq C$ implies that $D = D'$ and the set
\begin{align*}
  N(D/T_{\alpha+1}) = N(D'/T_\alpha)
\end{align*}
is linearly ordered. If, on the other hand, $D \subseteq C$ then we have
\begin{align*}
  N(D/T_{\alpha+1}) \subseteq N(C/T_\alpha) \cup \{v\}
\end{align*}
and the latter set is linearly ordered since $v$~is greater than every element
of $N(C/T_\alpha)$.
\qed
\begin{rem}\hfill
\begin{enuM}
\item(a)\enspace If the hypergraph~$\frakH$ is countable then we can actually
obtain a depth-first spanning tree of height at most~$\omega$ as follows.
In the above proof, if we are slightly more careful in choosing
the vertex~$v$ that is added to the partial tree,
then we can ensure that every vertex is chosen already after finitely many steps.

\item(b)\enspace Note that, strictly speaking, the above proposition is not a generalisation of
Theorem~1.4 of~\cite{Courcelle03} since we use a different notion
of a depth-first spanning tree.
\end{enuM}
\end{rem}

We use depth-first spanning trees to encode
orientations of a hypergraph.
First, we show that each depth-first spanning tree can be encoded
by finitely many $\GSO$-parameters.
\begin{lem}\label{Lem: d-f spanning trees definable}
For every $m < \omega$ we can construct $\MSO$-formulae
$\varphi(X;\bar Z)$, $\vartheta(x,Y;\bar Z)$, and $\chi(x,y;\bar Z)$
such that, for every connected hypergraph~$\frakH$
of rank at most~$m$ and each depth-first spanning tree $\langle T, {\leq}, (F_v)_v\rangle$ of~$\frakH$,
there are $\GSO$-parameters~$\bar S$ such that
\begin{alignat*}{-1}
  \frakH &\models \varphi(P;\bar S) &\quad\iff\quad
  &P \subseteq T \text{ is downward $\leq$-closed,} \\
  \frakH &\models \vartheta(v,P;\bar S) &\quad\iff\quad
  &v \in T \text{ and } P = A_v\,, \\
  \frakH &\models \chi(u,v;\bar S) &\quad\iff\quad
  &u,v \in T \text{ and } u \leq v\,.
\end{alignat*}
\end{lem}
\proof
We will use the following parameters\?:
\begin{enumerate}[$\bullet$]
\item Unary predicates $T_0$, $T_{s_l}$, $T_{t_l}$, for $l < m$,
  containing all vertices of the corresponding type.
\item $F^\tau := \bigcup {\set{ F_v }{ v \in T_\tau }}$, for every type~$\tau$.
\item $F := \bigcup_\tau F^\tau$\,.
\item $A^\tau := \bigcup {\set{ A_v }{ v \in T_\tau }} \setminus T$, for every type~$\tau$.
\item If $v$~is of type~$t_l$ then the set~$F_v$ forms a priority tree.
  We use additional parameters $F^{t_l}_i$ and $P_i$, for $i < m$,
  encoding the corresponding partition
  \begin{align*}
    F^{t_l} = F^{t_l}_0 \cupdot\dots\cupdot F^{t_l}_{m-1}
    \qtextq{and}
    \bigcup F^{t_l} = P_0 \cupdot\dots\cupdot P_{m-1}\,.
  \end{align*}
  Hence, for every vertex~$v$ of type~$t_l$ we have a priority tree
  \begin{align*}\textstyle
    \bigl\langle \bigcup F_v, F_v, L_v, (F^{t_l}_i \cap F_v)_i, (P_i \cap \bigcup F_v)_i, v\bigr\rangle\,.
  \end{align*}
\end{enumerate}
\begin{enuM}
\item(1)\enspace First, we construct the formula~$\vartheta$.
To simplify our task we define separate formulae
$\vartheta_\tau(x,Y)$, for each type~$\tau$, such that
\begin{align*}
  \frakH \models \vartheta_\tau(v,P) \quad\iff\quad v \in T_\tau \text{ and } P = A_v\,.
\end{align*}
Then we can set $\vartheta := \Lor_\tau\vartheta_\tau$.

If $v$~has type~$0$ then $A_v = \{v\}$ and we can set
\begin{align*}
  \vartheta_0(x,Y) := T_0x \land Y = \{x\}\,.
\end{align*}
If the type of~$v$ is~$s_l$ then $F_v = \{e\}$ and $A_v = \{v\} \cup (e \cap A^{s_l} )$,
where $e$~is the unique edge in $F^{s_l}$ containing~$v$.
Hence, we can define
\begin{align*}
  \vartheta_{s_l}(x,Y) := 
    T_{s_l} x
    \land (\exists e \in F^{s_l})[x \in e \land Y = \{x\} \cup (e \cap A^{s_l})]\,.
\end{align*}
Finally, if $v$~has type~$t_l$ then $A_v$ is the least subset of
$\{v\} \cup A^{t_l}$ satisfying the following conditions\?:
\begin{enumerate}[$\bullet$]
\item $v \in Y$
\item If $e \in F^{t_l}$ and $e \cap Y \neq \emptyset$ then $e \cap A^{t_l} \subseteq Y$.
\end{enumerate}
Hence, we can define
\begin{align*}
  \vartheta_{t_l}(x,Y) := {} &T_{t_l}x
  \land Yx \land (\forall e \in F^{t_l})(e \cap Y \neq \emptyset \lso e \cap A^{t_l} \subseteq Y) \\
  {} \land {}&
    \forall Z[Zx \land (\forall e \in F^{t_l})(e \cap Z \neq \emptyset \lso e \cap A^{t_l} \subseteq Z)
              \lso Y \subseteq Z]\,.
\end{align*}

\item(2)\enspace Next, we define a formula~$\alpha(x,Y)$ such that
\begin{align*}
  \frakH \models \alpha(v,Q)
  \quad\iff\quad
  v \in T \text{ and }\textstyle
  Q = B(\bigcup F_v/T)\,.
\end{align*}
The formula $\alpha(x,Y)$ should state that
\begin{align*}
  Y = \set{ u \in T }{ \text{there is some } e \in F_v \text{ with } e \cap A_x \neq \emptyset
    \text{ and } e \cap A_u \neq \emptyset }\,.
\end{align*}
Using the formulae~$\vartheta_\tau$ we can write~$\alpha$ as
\begin{align*}
  \alpha(x,Y) :=
    \Lor_\tau [T_\tau x \land (\exists e \in F^\tau)(e \cap A_x \neq \emptyset \land e \cap A_u \neq \emptyset)]\,.
\end{align*}

\item(3)\enspace With the help of~$\alpha$ we can write down the desired formulae $\varphi$~and~$\chi$.
\begin{align*}
  \varphi(X) &:= \forall x\forall Y[Xx \land \alpha(x,Y) \lso Y \subseteq X]\,, \\
  \chi(x,y)  &:= \forall X[\varphi(X) \lso (Xy \lso Xx)]\,.
\end{align*}
\upqed
\qed
\end{enuM}

\begin{lem}\label{Lem: preorder on auxiliary sets}
We can construct\/ $\GSO$-formulae $\psi_m(x,y;\bar Z)$, for $m < \omega$, such that,
for every depth-first spanning tree $\langle T, {\leq}, (F_v)_v\rangle$
of a connected hypergraph\/~$\frakH$ of rank~$m$, there are\/ $\GSO$-parameters~$\bar S$ such that
the formula $\psi_m(x,y;\bar S)$ defines a preorder\/~$\sqsubseteq_0$ with the
following properties\?:
\begin{enumerate}[$\bullet$]
\item The restriction of\/~$\sqsubseteq_0$ to~$T$ coincides with~$\leq$.
\item $\sqsubseteq_0$~linearly preorderes every set $X \subseteq V$ such that
  $B(X/T)$ is linearly ordered by~$\leq$.
\item Each\/ $\sqsubseteq_0$-class has at most~$m$ elements.
\end{enumerate}
\end{lem}
\proof
Let $\chi(x,y)$~and~$\vartheta(x,Y)$ be the formulae of
Lemma~\ref{Lem: d-f spanning trees definable}.
For each type~$\tau$, we define a formula~$\eta_\tau(x,y)$ linearly preordering
each set~$A_v$ where $v$~is of type~$\tau$.
Then the desired formula~$\psi_m$ states that either
\begin{enumerate}[(1)]
\item $x \in A_u$ and $y \in A_v$ for $u < v$, or
\item $x,y \in A_v$, for some~$v$ of type~$\tau$, and $\eta_\tau(x,y)$ holds.
\end{enumerate}
If $v$~is of type $0$~or~$s_l$ then $A_v$ contains at most~$m$
elements and we can set
\begin{align*}
  \eta_\tau(x,y) := \mathrm{true}\,.
\end{align*}
For vertices of type~$t_l$ we can use the formula from
Lemma~\ref{Lem: existence of priority trees}.
\qed
\begin{cor}\label{Cor: order on auxiliary sets}
We can construct\/ $\GSO$-formulae $\psi_m(x,y;\bar Z)$, for $m < \omega$, such that,
for every depth-first spanning tree $\langle T, {\leq}, (F_v)_v\rangle$
of a connected hypergraph~$\frakH$ of rank~$m$, there are $\GSO$-parameters~$\bar S$ such that
the formula $\psi_m(x,y;\bar S)$ defines a partial order~$\sqsubseteq$ with the
following properties\?:
\begin{enumerate}[$\bullet$]
\item The restriction of\/~$\sqsubseteq$ to~$T$ coincides with~$\leq$.
\item $\sqsubseteq$~linearly orderes every set $X \subseteq V$ such that
  $B(X/T)$ is linearly ordered by~$\leq$.
\end{enumerate}
\end{cor}
\proof
Let $\sqsubseteq_0$~be the preorder from Lemma~\ref{Lem: preorder on auxiliary sets}.
Since every $\sqsubseteq_0$-class contains at most~$m$ elements we can add $m$~new unary predicates
$P_0,\dots,P_{m-1}$ such that $P_0 \cup\dots\cup P_{m-1} = V$ and
we have $\abs{X \cap P_i} \leq 1$, for each $\sqsubseteq_0$-class~$X$ and all~$i$.
Then we can define
\begin{align*}
  u \sqsubseteq v \quad\defiff\quad
  &\text{either } u \sqsubset_0 v\,, \text{ or} \\
  &u \sqsubseteq_0 v\,,\ v \sqsubseteq_0 u\,,\ u \in P_i\,,\ v \in P_k \text{ for } i < k\,.
\end{align*}
\upqed
\qed

\begin{thm}
We can construct\/ $\GSO$-formulae $\varphi_m(x,Y;\bar Z)$, for $m < \omega$, such that
for every hypergraph~$\frakH$ of rank~$m$, there are $\GSO$-parameters~$\bar S$ such that,
the formula $\varphi_m(x,Y;\bar S)$ defines an orientation of~$\frakH$.
\end{thm}
\proof
Suppose that $\frakH$~has $\kappa$~connected components~$C_i$, $i < \kappa$.
For each component~$C_i$ we
fix a depth-first spanning tree $\langle T^i, {\leq^i}, (F^i_v)_v\rangle$.
Let $\bar S^i$~be the parameters from Lemma~\ref{Lem: d-f spanning trees definable}
and Lemma~\ref{Lem: preorder on auxiliary sets}.
For every edge~$e \in E$, there exists a unique component~$C_i$
such that the intersection $X := e \cap \bigcup_v A_v^i$ is finite and nonempty.
Furthermore, the set $B(X/T^i)$ is linearly ordered by~$\leq$.
Using the ordering~$\sqsubseteq$ of Corollary~\ref{Cor: order on auxiliary sets}
we can write down a formula $\varphi_m(v,e)$ stating that
$v$~is the $\sqsubseteq$-least element of this set~$X$.
\qed
\begin{cor}
Every hypergraph of rank $m < \omega$ is $\GSO$-orientable.
\end{cor}

Let us mention the following consequences of this result.
For countable hypergraphs they are again due to Courcelle~\cite{Courcelle03}.
\begin{defi}\hfill
\begin{enuma}
\item A formula $\varphi(x,y,Z)$ defines an \emph{edge ordering} of a hypergraph
$\frakH = \langle V,E\rangle$
if, for every edge $e \in E$, the formula $\varphi(x,y,e)$ defines a linear
ordering on the vertices of~$e$.

\item A formula $\varphi(x,y,z)$ defines an \emph{neighbourhood ordering} of
a directed graph $\frakG = \langle V,E\rangle$
if, for every vertex $v \in V$, the formula $\varphi(x,y,v)$ defines a
linear ordering on the set
$\set{ u \in V }{ (u,v) \in E }$.
\end{enuma}
\end{defi}

\begin{lem}\label{Lem: edge ordering}
There exist $\GSO$-formulae $\varphi_m(x,y,Z;\bar U)$, for $m < \omega$,
such that, for every hypergraph $\frakH = \langle V,E\rangle$ of rank~$m$,
there are $\GSO$-parameters~$\bar S$
such that the formula $\varphi_m(x,y,Z;\bar S)$ defines an edge ordering of~$\frakH$.
\end{lem}
\begin{lem}\label{Lem: neighbourhood ordering}
There exist $\MSO$-formulae $\varphi_m(x,y,z;\bar U)$, for $m < \omega$,
such that, for every directed graph~$\frakG$ of indegree at most~$m$,
there are $\MSO$-parameters~$\bar P$
such that the formula $\varphi_m(x,y,z;\bar P)$ defines a neighbourhood ordering of~$\frakG$.
\end{lem}
\proof
We can apply Lemma~\ref{Lem: edge ordering} to the hypergraph $\frakH := \langle V,F\rangle$
where
\begin{align*}
  F := \set{ I(v) }{ v \in V } \qtextq{with} I(v) := \set{ u \in V }{ (u,v) \in E }\,.
\end{align*}
Note that every subset $S \subseteq F$ can be encoded by the set
\begin{align*}
  I^{-1}(F) := \set{ v \in V }{ I(v) \in F } \subseteq V\,.
\end{align*}
Hence, every $\GSO$-formula over~$\frakH$ can be translated into an $\MSO$-formula
over~$\frakG$.
\qed

\section{\texorpdfstring{$\GSO$}{GSO} versus \texorpdfstring{$\MSO$}{MSO}}
\label{Sect: GSO and MSO}

\noindent In \cite{Courcelle03} Courcelle has shown that
we can translate every $\GSO$-formula~$\varphi$ into an $\MSO$-formula~$\psi$
that is equivalent to~$\varphi$ on all countable $k$-sparse hypergraphs.
Using the results of the previous sections we can lift the restriction to countable
hypergraphs.
The proof in~\cite{Courcelle03} goes through unchanged since it relies
only on the statements of Lemma~\ref{Lem: edge ordering} and Lemma~\ref{Lem: neighbourhood ordering},
and on local modifications of hypergraphs.
\begin{thm}
For all numbers $m,k < \omega$, there
exists a monadic second-order interpretation (with monadic parameters)
that maps a $k$-sparse hypergraph of rank~$m$ to its incidence structure.
\end{thm}
\begin{cor}
For all $m,k < \omega$ and all formulae $\varphi(\bar x,\bar Y,\bar Z) \in \GSO$
with first-order variables~$\bar x$, monadic variables~$\bar Y$,
and guarded second-order variables~$\bar Z$,
there exists a formula $\psi(\bar x,\bar Y,\bar Z) \in \MSO$ with the following property\?:
for all $k$-sparse hypergraphs $\frakH = \langle V,E\rangle$ of rank~$m$ and all parameters
$a_i \in V$, $P_i \subseteq V$, $R_i \subseteq E$,
there exist parameters $Q_i \subseteq V$ such that
\begin{align*}
  \frakH \models \varphi(\bar a,\bar P,\bar R)
  \quad\iff\quad
  \frakH \models \psi(\bar a,\bar P,\bar Q)\,.
\end{align*}
\end{cor}

\section{Sparse distributions}   
\label{Sect: flow}

\noindent The results so far concern ways to encode edges by vertices.
In this last section we consider a more general problem.
Let $\frakG = \langle V,E\rangle$ be a graph.
We denote by $\PSet_{\mathrm{fin}}(V)$ the set of all finite subsets of~$V$.
We would like to encode a given subset $F \subseteq \PSet_{\mathrm{fin}}(V)$
by a set of vertices, that is, we would like to find a definable
function $h : F \to V$ that is injective.
For $F = E$ this reduces to the problem considered in the preceding sections.
For arbitrary~$F$, such a function~$h$ does not always exist.
But we will show that sometimes we can transform a given function $h_0 : F \to V$
into an injective one.

These results are inspired by work of
Colcombet and L\"oding~\cite{ColcombetLoeding07} on set interpretations.
Colcombet and L\"oding consider a power set operation~$\calP$ on
structures. One of their main results in a commutation theorem for
interpretations and the power set operation.
They show that, given a tree~$\frakT$ and
an $\FO$-interpretation~$\calI$ such that $\calI(\calP(\frakT))$
is of the form $\calP(\frakM)$, for some structure~$\frakM$, then there
exists a $\WMSO$-inter\-pre\-ta\-tion~$\calJ$ such that $\frakM \cong \calJ(\frakT)$.
On ingredient in the proof of this result is a method to encode,
in a definable way, finite subsets of the tree~$\frakT$ by single vertices.

Suppose we are given a function $h_0 : F \to V$ that we want to transform
into an injective function $h : F \to V$.
Let $\delta(v) := \abs{h_0^{-1}(v)}$.
The first step in the construction of~$h$ consists in finding
a definable function $g : V \to V$
such that $\abs{g^{-1}(v)} = \delta(v)$, for all~$v$.
Of course, this is not always possible. For instance, if the graph is
finite and we have $\delta(v) > 1$, for all vertices~$v$.
Therefore, we consider only functions~$\delta$ that are \emph{sparse}
in the sense of the following definition.
\begin{defi}
Let $\frakG = \langle V,E\rangle$ be an undirected graph.
\begin{enuma}
\item The \emph{border} of a subset $Z \subseteq V$
is the set
\begin{align*}
  B_\frakG(Z) := E \cap (V \setminus Z) \times Z
\end{align*}
of all edges connecting a vertex in~$Z$ with a vertex outside of~$Z$.

\item A \emph{distribution} of~$\frakG$ is a map
$\delta : V \to \omega$. For $X \subseteq V$, we define the shorthand
\begin{align*}
  \delta(X) := \sum_{v \in X} \delta(x)\,.
\end{align*}

\item Let $h : X \to V$ be an arbitrary mapping.
The distribution \emph{induced} by~$h$ is the function $\delta : V \to \omega$ with
\begin{align*}
  \delta(v) := \abs{h^{-1}(v)}\,.
\end{align*}

\item A distribution~$\delta$ is \emph{$k$-sparse} if
\begin{align*}
  \delta(Z) \leq \abs{Z} + k\cdot\abs{B_\frakG(Z)}\,,
  \quad\text{for every } Z \subseteq V\,.
\end{align*}
\end{enuma}
\end{defi}

Given a $k$-sparse distribution~$\delta$ we will construct the
desired function $g : V \to V$ by solving a network flow problem.
\begin{defi}
Let $\frakG = \langle V,E\rangle$ be an undirected graph.
\begin{enuma}
\item A \emph{flow} of~$\frakG$ is a function $f : V \times V \to \bbZ$
such that, for all $u,v \in V$,
\begin{enumerate}[$\bullet$]
\item $f(u,v) = -f(v,u)$ and
\item $f(u,v) \neq 0$ implies $(u,v) \in E$.
\end{enumerate}

\item A flow~$f$ is \emph{acyclic} if there is no cycle $u_0,\dots,u_m$ of~$\frakG$
such that $f(u_m,u_0) > 0$ and $f(u_i,u_{i+1}) > 0$, for all $i < m$.

\item The \emph{defect} of a flow~$f$ is the distribution
\begin{align*}
  d_f(v) := \sum_{u \in V} f(v,u)\,.
\end{align*}

\item A flow~$f$ is a \emph{$\delta$-flow} if, for every $v \in V$, either
\begin{align*}
  d_f(v) = \delta(v) - 1
  \quad\text{or}\quad
  \delta(v) = 0 \text{ and\/ } d_f(v) = 0\,.
\end{align*}

\item A flow~$f$ is \emph{edge-bounded} by~$k$ if $\abs{f(u,v)} \leq k$, for all $u,v \in V$.
We call~$f$ \emph{vertex-bounded} by~$k$ if
\begin{align*}
  \sum_{u \in V} \abs{f(u,v)} \leq k\,, \quad\text{for all } v \in V\,.
\end{align*}
\end{enuma}
\end{defi}

\noindent Our aim is to show that, for every $k$-sparse distribution~$\delta$
there is a bounded $\delta$-flow~$f$ and a function $g : V \to V$ inducing~$\delta$.
Furthermore, if $\delta$~is definable then $g$~should also be definable.
\begin{defi}
Let $L$~be a logic.
\begin{enuma}
\item A distribution~$\delta$ is \emph{$L$-definable}
if there exist formulae $\varphi_i(x) \in L$, $i < k$, such that
\begin{align*}
  \frakG \models \varphi_i(v) \quad\iff\quad \delta(v) = i\,.
\end{align*}

\item Similarly, a flow~$f$ is \emph{$L$-definable} if there exist
formulae $\varphi_i(x,y) \in L$ such that
\begin{align*}
  \frakG \models \varphi_i(u,v) \quad\iff\quad f(u,v) = i\,.
\end{align*}
\end{enuma}
\end{defi}
\begin{rem}
Note that every edge-bounded flow can be encoded with the help of
the $\GSO$-parameters
\begin{align*}
  S_i := \set{ (u,v) \in E }{ f(u,v) = i }\,.
\end{align*}
\end{rem}

For trees the problem of encoding sets by vertices
has been solved by Colcombet and L\"oding~\cite{ColcombetLoeding07}.
In the general case proved below the function~$g$
is only definable with the help of $\GSO$-parameters, but for trees
we can do without them.

\begin{thm}[Colcombet and L\"oding~\cite{ColcombetLoeding07}]
Let\/ $\frakT = \langle T,E\rangle$ be an infinite directed tree
and $\delta$~a\/ $\WMSO$-definable $k$-sparse distribution of\/~$\frakT$.
There exists a\/ $\WMSO$-definable flow~$f$ that is
edge-bounded by~$7k$ and satisfies $d_f(v) \geq \delta(v) - 1$, for all~$v$.
\end{thm}

\begin{thm}[Colcombet and L\"oding~\cite{ColcombetLoeding07}]
Let\/ $\frakT = \langle T,E\rangle$ be a directed tree
and $\delta$~a\/ $\WMSO$-definable $k$-sparse distribution of\/~$\frakT$
such that $\delta(T) \leq \abs{T}$.
There exists\/ $\WMSO$-definable function $g : T \to T$ such that $\delta$~is
the distribution induced by~$g$.
\end{thm}

To prove our generalisation of these results
we start with a few lemmas about bounded flows.
The first two follow immediately from the definitions.
\begin{lem}
Every flow that is vertex-bounded by~$k$ is also edge-bounded by~$k$.
\end{lem}
\begin{lem}
Suppose that\/ $\frakG$~is a graph with maximal degree~$d$.
Every flow of\/~$\frakG$ that is edge-bounded by~$k$ is
vertex-bounded by~$dk$.
\end{lem}

\begin{lem}
For every $\delta$-flow~$f$ there exists an acyclic $\delta$-flow~$f'$ such that,
if $f$~is edge-bounded by~$k$ or vertex-bounded by~$k$ then so is~$f'$.
\end{lem}
\proof
We repeat the following construction until the flow is acyclic.
Select a cycle $u_0,\dots,u_m$ such
that
\begin{align*}
  c := \min {\set{ f(u_i,u_{i+1}) }{ i \leq m }} > 0\,.
\end{align*}
We define $f'$~by
\begin{align*}
  f'(x,y) := \begin{cases}
               f(x,y)-c &\text{if } x = u_i \text{ and } y = u_{i+1}\,, \text{ for some } i\,, \\
               f(x,y)+c &\text{if } x = u_{i+1} \text{ and } y = u_i\,, \text{ for some } i\,, \\
               f(x,y)   &\text{otherwise}\,.
             \end{cases}
\end{align*}
\upqed
\qed

\begin{prop}\label{Prop: existence of flows}
Let $\frakG = \langle V,E\rangle$ be an undirected graph and $\delta$~a $k$-sparse distribution.
Then $\frakG$~has a $\delta$-flow~$f$ that is edge-bounded by~$k$.
\end{prop}
\proof
First, we assume that $\frakG$~is finite. In this case we can reduce the task to
a network flow problem.
Let $\frakH$~be the graph obtained from~$\frakG$ by adding two new vertices $s$~and~$t$
that are connected to every vertex of~$\frakG$. We define the capacity~$c(e)$
of edges~$e$ of~$\frakH$ as follows.
For edges~$e$ of~$\frakG$ we set $c(e) := k$.
If $e = (s,v)$ with $v \in V$ we set $c(e) := \max \{0,\ \delta(v) - 1\}$.
Finally, if $e = (v,t)$ with $v \in V$ we define
\begin{align*}
  c(e) := \begin{cases}
            0 &\text{if } \delta(v) > 0\,, \\
            1 &\text{otherwise}\,.
          \end{cases}
\end{align*}
Let $f$~be a maximal flow from~$s$ to~$t$ with respect to~$c$.
We claim that its restriction to the edges of~$\frakG$ is the desired flow.

According to the Max-Flow Min-Cut Theorem, there is a set~$X$ of vertices containing~$s$
but not~$t$ such that the maximal flow~$m$ from~$s$ to~$t$ equals
\begin{align*}
  m = \sum_{e \in B_\frakH(X)} c(e)\,.
\end{align*}
Let $X_0 := X \setminus \{s\} \subseteq V$ and $Y := \delta^{-1}(0)$. Since
\begin{align*}
  B_\frakH(X) = B_\frakG(X_0) \cup \set{ (v,t) }{ v \in X_0 } \cup \set{ (s,v) }{ v \in V \setminus X_0 }\,,
\end{align*}
we have
\begin{align*}
  m &= \sum_{e \in B_\frakH(X)} c(e) \\
    &= k\cdot \abs{B_\frakG(X_0)} + \abs{X_0 \cap Y}
       + \delta(V \setminus X_0) - \abs{(V \setminus X_0) \setminus Y} \\
    &= k\cdot \abs{B_\frakG(X_0)} + \abs{X_0} + \delta(V \setminus X_0)
       - \abs{(V \setminus X_0) \setminus Y} - \abs{X_0 \setminus Y} \\
    &\geq \delta(X_0) + \delta(V \setminus X_0) - \abs{V \setminus Y} \\
    &= \delta(V) - \abs{V \setminus Y}\,.
\end{align*}
On the other hand, for the set $X = \{s\}$, we have
\begin{align*}
  m \leq \sum_{e \in B_\frakH(X)} c(e) = \sum_{v \in V} \max \{0,\ \delta(v)-1\}
     = \delta(V) - \abs{V \setminus Y}\,.
\end{align*}
Consequently, the maximal flow~$m$ from~$s$ to~$t$ equals
\begin{align*}
  m = \delta(V) - \abs{V \setminus Y}\,.
\end{align*}
This implies that
\begin{align*}
  f(s,v) = \max \{0,\ \delta(v)-1\}\,, \quad\text{for every } v \in V\,.
\end{align*}
For each $v \in V$, we therefore have
\begin{align*}
  0 = \sum_{u \in V \cup \{s,t\}} f(u,v)
    = \max \{0,\delta(v)-1\} + f(t,v) + \sum_{u \in V} f(u,v)\,.
\end{align*}
If $\delta(v) > 0$ this implies
\begin{align*}
  \delta(v)-1 - \sum_{u \in V} f(v,u) = 0\,,
  \qtextq{that is}
  d_f(v) = \delta(v) - 1\,,
\end{align*}
while, for $\delta(v) = 0$, we have
\begin{align*}
  {-f(v,t)} - \sum_{u \in V} f(v,u) = 0\,.
\end{align*}
Hence, either $d_f(v) = {-1} = \delta(v) - 1$ or $d_f(v) = 0$.

It remains to prove the lemma for infinite graphs. Let $\Phi(\frakG)$
consist of the elementary diagram of~$\frakG$
together with first-order formulae stating that $f$~is a $\delta$-flow
on~$\frakG$ that is edge-bounded by~$k$.
We will use the compactness theorem to show that $\Phi(\frakG)$ is satisfiable.

Let $\Phi_0 \subseteq \Phi(\frakG)$ be finite. There exists a finite induced
subgraph $\frakG_0 = \langle V_0,E_0\rangle$ of~$\frakG$ such that
$\Phi_0 \subseteq \Phi(\frakG_0)$. Let $\langle u_0,v_0\rangle,\dots,\langle u_{m-1},v_{m-1}\rangle$
be an enumeration (without repetitions) of all edges $\langle u,v\rangle$ with $u \in V_0$ and
$v \in V \setminus V_0$. We construct a new graph $\frakG_0' = \langle V_0',E_0'\rangle$ by attaching
to each vertex~$u_i$ a path~$P_i$ of length~$k$. Let $\delta'$~be the distribution on~$\frakG_0'$
with $\delta'(v) = \delta(v)$, for $v \in V_0$, and $\delta'(v) = 0$, for $v \in V_0' \setminus V_0$.
In order to show that $\Phi_0$~is satisfiable it is sufficient to prove that $\frakG_0'$~has a flow
of the desired form. Consider an arbitrary set $X \subseteq V_0'$ of vertices. Let
\begin{align*}
  I := \set{ i }{ u_i \in X } \quad\text{and}\quad
  J := \set{ i }{ u_i \in X \text{ and } P_i \subseteq X }\,.
\end{align*}
It follows that
\begin{align*}
  \delta'(X)
  = \delta(X \cap V_0)
  &\leq \abs{X \cap V_0} + k\cdot\abs{B_\frakG(X \cap V_0)} \\
  &\leq \abs{X} - k\cdot\abs{J} + k\cdot\abs{B_{\frakG_0'}(X \cap V_0)} \\
  &\leq \abs{X} - k\cdot\abs{J} + k\cdot\bigl(\abs{B_{\frakG_0'}(X)} + \abs{J}\bigr)
   = \abs{X} + k\cdot\abs{B_{\frakG_0'}(X)}\,.
\end{align*}
By the first part of the proof it follows that $\frakG_0'$~has a flow of the desired form.
\qed

It remains to show how we can use the $\delta$-flow~$f$ we have just constructed
to define the desired function $g : V \to V$.
We start by selecting a certain family of definable paths.
Note that we allow paths of length~$0$. Such paths are
uniquely determined by the vertex they start (and end) at.
\begin{lem}\label{Lem: paths coding a flow}
Let $\frakG$~be a countable undirected graph and $f$~an acyclic $\delta$-flow of~$\frakG$.
There exists a set~$\calP$ of finite paths through~$\frakG$
satisfying the following conditions\?:
\begin{enumi}
\item For every $v \in V$, there are exactly $\delta(v)$
  paths in~$\calP$ starting at~$v$.
\item For every $v \in V$ there is at most one path in~$\calP$ ending at~$v$.
\item For every pair $u,v \in V$ of vertices there are at most $f(u,v)$ paths
  in~$\calP$ containing the edge $(u,v)$ (in this direction).
\end{enumi}
\end{lem}
\proof
Fix an enumeration $(v_n,k_n)_{n < \omega}$ of the set
\begin{align*}
  \set{ \langle v,k\rangle }{ v \in V,\ 0 \leq k < \delta(v) }\,.
\end{align*}
For $n < \omega$, we construct paths~$\pi_n$ with the following properties\?:
\begin{enumerate}[$\bullet$]
\item $\pi_n$~starts at~$v_n$.
\item If $m \neq n$ then the endpoints of $\pi_m$~and~$\pi_n$ are different.
\item For every edge $(u,v)$ there are at most $f(u,v)$ paths~$\pi_n$ containing the edge $(u,v)$.
\end{enumerate}
By induction, suppose that we have already defined~$\pi_i$, for $i < n$. Let
\begin{enumerate}[(1)]
\item $\alpha(v)$ be the number of paths~$\pi_i$, $i < n$, starting at~$v$,
\item $\beta(v)$ the number of paths~$\pi_i$, $i < n$, ending at~$v$, and
\item $\mu(u,v)$ the number of paths~$\pi_i$, $i < n$, containing the edge $(u,v)$.
\end{enumerate}
We construct a path $u_0\dots u_m$ inductively starting with $u_0 := v_n$.
For the induction step, suppose that we have already defined $u_0,\dots,u_i$.
If $\beta(u_i) = 0$ then we stop and set $\pi_n := u_0\dots u_i$.
Otherwise, we claim that there is some neighbour~$w$ of~$u_i$ with
$f(u_i,w) > \mu(u_i,w)$. Hence, we can set $u_{i+1} := w$.

To prove the claim, we distinguish two cases. If $i = 0$ then $\alpha(u_0) < \delta(u_0)$
implies that
\begin{align*}
  \sum_{x \in V} \mu(u_0,x)
    &= \alpha(u_0) - \beta(u_0) + \sum_{x \in V} \mu(x,u_0) \\
    &\leq \alpha(u_0) - 1 + \sum {\set{ f(x,u_0) }{ f(x,u_0) \geq 0 }} \\
    &= \alpha(u_0) - 1 + \sum {\set{ f(u_0,x) }{ f(u_0,x) \geq 0 }} - (\delta(u_0) - 1) \\
    &< \sum {\set{ f(u_0,x) }{ f(u_0,x) \geq 0 }}\,,
\end{align*}
as desired. Similarly, if $i > 0$ then
$\mu(u_{i-1},u_i) < f(u_{i-1},u_i)$ implies that
\begin{align*}
  \sum_{x \in V} \mu(u_i,x)
    &= \alpha(u_i) - \beta(u_i) + \sum_{x \in V} \mu(x,u_i) \\
    &< \alpha(u_i) - 1 + \sum {\set{ f(x,u_i) }{ f(x,u_i) \geq 0 }} \\
    &= \alpha(u_i) - 1 + \sum {\set{ f(u_i,x) }{ f(u_i,x) \geq 0 }} - (\delta(u_i) - 1) \\
    &\leq \sum {\set{ f(u_i,x) }{ f(u_i,x) \geq 0 }}\,.
\end{align*}
Note that the construction of~$\pi_n$ must terminate after at most $n+1$ steps
since the flow~$f$ is acyclic and there are only~$n$ vertices~$u$ with $\beta(u) = 1$.
\qed

\begin{lem}\label{Lem: encoding a set of paths}
There exist $\GSO$-formulae $\varphi_m(X;\bar Z)$, for $m < \omega$,
such that, for every graph\/~$\frakG$ and each set~$\calP$ of finite paths
such that every vertex and every edge of\/~$\frakG$ is contained in at most~$m$ paths
of~$\calP$, there exists a tuple~$\bar S$ of $\GSO$-parameters such that
\begin{align*}
  \frakG \models \varphi_m(P;\bar S)
  \quad\iff\quad
  P \text{ is \textup(the set of edges of\textup) a nonempty path in } \calP\,.
\end{align*}
\end{lem}
\proof
For every edge $(u,v)$ of~$\frakG$ we fix a bijection
$\mu(u,v) : [n] \to \calP_e$ where $\calP_e \subseteq \calP$
is the set of all paths containing the edge $(u,v)$ (in either direction)
and $n := \abs{\calP_e}$. We assume that $\mu(u,v) = \mu(v,u)$.

Let $S$~be the set of all edges of~$\frakG$ contained in some path in~$\calP$.
By Lemma~\ref{Lem: neighbourhood ordering} there exists an
$\MSO$-formula $\chi(x,y,z;\bar S')$ with parameters~$\bar S'$
such that, for every $v \in V$, the formula $\chi(x,y,v;\bar S')$ linearly orders the set of
all vertices that are connected to~$v$ via an edge in~$S$.

Finally, we define unary predicates~$Q^{ik}_{jl}$ containing all vertices~$v$
such that there exists a path $\pi \in \calP$ containing edges $(u,v)$, $(v,w)$
where
\begin{enumerate}[$\bullet$]
\item $\mu(u,v)(k) = \pi$, $\mu(v,w)(l) = \pi$,
\item $u$~is the $i$-th neighbour of~$v$ (in the order defined by~$\chi$),
\item $w$~is the $j$-th neighbour of~$v$.
\end{enumerate}
It follows that a nonempty set $P \subseteq E$ of edges is a path in~$\calP$ if
and only if
$P$~is a minimal nonempty subset of~$E$ satisfying the following condition\?:
\begin{enumerate}[$-$]
\item[] $P$~can be written as a union $P = P_0 \cup\dots\cup P_{m-1}$ such that,
  for all vertices $u,v,w$ such that
  $v \in Q^{ik}_{jl}$ and $u$~and~$w$ are, respectively, the $i$-th and $j$-th neighbour of~$v$,
  we have $(u,v) \in P_k \Leftrightarrow (v,w) \in P_l$.
\end{enumerate}
This condition can be expressed in~$\GSO$.
\qed
\begin{rem}
Note that the set of empty paths in~$\calP$ is trivially definable
with the help of the parameter
\begin{align*}
  Q := \set{ v \in V }{ \calP \text{ contains an empty path from $v$ to } v }\,.
\end{align*}
\end{rem}

Using the family~$\calP$ we can construct a formula~$\varphi$ defining the function~$g$.
\begin{prop}\label{Prop: function from a flow}
There exist $\GSO$-formulae $\varphi_m(x,y;\bar Z)$, for $m < \omega$,
with the following property\?:
for every graph $\frakG = \langle V,E\rangle$ and each acyclic $\delta$-flow~$f$ of~$\frakG$
that is vertex-bounded by~$m$,
there exist $\GSO$-parameters~$\bar S$ such that
$\varphi_m(x,y;\bar S)$ defines on~$\frakG$ a partial function $g : V \to V$ with
\begin{align*}
  \abs{g^{-1}(v)} = \delta(v)\,, \quad\text{for all } v \in V\,.
\end{align*}
\end{prop}
\proof
Let $\frakG'$~be the graph obtained from~$\frakG$ by removing every edge
$(u,v)$ with $f(u,v) = 0$. Note that $f$~is also a $\delta$-flow of~$\frakG'$.
Since $f$~is vertex-bounded by~$m$ it follows that
every vertex of~$\frakG'$ has degree at most~$m < \omega$.
Consequently, each connected component~$\frakG_0$ of~$\frakG$ is countable.
Let $\calP_0$~be the set of paths obtained by applying Lemma~\ref{Lem: paths coding a flow}
to the restriction of~$f$ to~$\frakG_0$, and let $\calP$~be the union of all
these sets~$\calP_0$ corresponding to the connected components of~$\frakG'$.
By Lemma~\ref{Lem: encoding a set of paths}, there exists a formula
$\psi(X;\bar Z)$ and a set~$\bar S$ of guarded relations such that
\begin{align*}
  \frakG \models \psi(P;\bar S)
  \quad\iff\quad P \text{ is a nonempty path in } \calP\,.
\end{align*}
With the help of~$\psi$ we can define
a partial function $g : V \to V$ such that
\begin{align*}
  g(v) = u \quad\defiff\quad \calP \text{ contains a path from } u \text{ to } v\,.
\end{align*}
By construction of~$\calP$ we have $\abs{g^{-1}(v)} = \delta(v)$, for every $v \in V$.
\qed

\begin{lem}
Let $\frakG = \langle V,E\rangle$ be a graph of finite degree and
$\varphi(X,y)$ a $\GSO$-formula that defines a partial function $h : \PSet(V) \to V$
such that the distribution~$\delta$ induced by~$h$ is $k$-sparse.
Suppose that there exists a $\GSO$-formula $\chi(X,Y,z)$ such that,
for every vertex $v \in V$, $\chi(X,Y,v)$ linearly orders the set
$h^{-1}(v)$.
Then there exist $\MSO$-definable partial functions $h_0 : \PSet(V) \to V$ and $g : V \to V$
such that $h = g \circ h_0$ and $h_0$~is injective.
\end{lem}
\proof
By Proposition~\ref{Prop: existence of flows}
there exists a $\delta$-flow~$f$ that is edge-bounded by~$k$.
Since $\frakG$~has finite degree it follows that $f$~is vertex-bounded
by some constant $m < \omega$.
Hence, we can use Proposition~\ref{Prop: function from a flow}
to find a definable function $g : V \to V$ with
$\abs{g^{-1}(v)} = \delta(v) = \abs{h^{-1}(v)}$.
Choose unary predicates $P_0,\dots,P_{k-1}$ such that we have
$i \neq l$ whenever
$u \in P_i$ and $v \in P_l$ are distinct vertices with
$g(u) = g(v)$.
Using these predicate we can define partial functions $g_0,\dots,g_{k-1} : V \to V$
such that $g_i(v)$~is the unique element of $g^{-1}(v) \cap P_i$.
We define $h_0 : \PSet(V) \to V$ by $h_0(X) := (g_i \circ h)(X)$
where the index~$i$ is chosen such that
$X$~is the $i$-th element of $h^{-1}(h(X))$ (in the order defined by~$\chi$).
It follows that $h(X) = g(h_0(X))$ and $h_0$~is injective.
Furthermore, the function~$h_0$ is clearly $\GSO$-definable. Since the graph~$\frakG$
has degree at most~$k$ it is $k$-sparse. Hence, every $\GSO$-definable
function is already $\MSO$-definable.
\qed

Recall that $\PSet_{\mathrm{fin}}(V)$ denotes the set of all finite subsets of~$V$.
Combining the preceding lemmas we obtain the main result of this section.
\begin{thm}\label{Thm: coding sets by vertices}
Let $\frakG = \langle V,E\rangle$ be a graph of finite degree and
$\varphi(X,y)$ a $\GSO$-formula that defines a partial function $h : \PSet_{\mathrm{fin}}(V) \to V$
such that the distribution~$\delta$ induced by~$h$ is $k$-sparse.
Then there exist $\MSO$-definable partial functions $h_0 : \PSet_{\mathrm{fin}}(V) \to V$ and $g : V \to V$
such that $h = g \circ h_0$ and $h_0$~is injective.
\end{thm}
\proof
By the preceding lemma it is sufficient to construct a formua~$\chi(X,Y,z)$
(with $\GSO$-parameters) such that $\chi(X,Y,v)$ linearly orders $h^{-1}(v)$,
for every $v \in V$.
Let $T_0 \subseteq E$ be a spanning forest of~$\frakG$
and let $P \subseteq V$ be a set containing exactly one element of each connected component.
Using the parameters $P$~and~$T_0$ we can define the tree ordering on~$V$ by
\begin{align*}
  u \leq v \quad\defiff\quad
    \text{the unique path in } T_0 \text{ from some element of } P \text{ to } v
    \text{ contains } u\,.
\end{align*}
Let $T \subseteq V \times V$ be the set obtained from~$T_0$
by orienting the edges according to this ordering.
Then $T$~is a directed forest. Furthermore, since the degree of~$\frakG$ is bounded
we can use Lemma~\ref{Lem: neighbourhood ordering} to linearly order the successors
of every vertex in~$T$. We use these two orderings to define the lexicographic
ordering~$\leq_\lex$ on~$T$.
Finally, we obtain the desired ordering on $\PSet_{\mathrm{fin}}(V)$ by setting
\begin{align*}
  X < Y \quad\defiff\quad
    \text{the $\leq_{\mathrm{lex}}$-minimal element of }
    (X \setminus Y) \cup (Y \setminus X) \text{ belongs to } Y\,.
\end{align*}
Each of these definitions can be expressed in~$\GSO$.
\qed

\section{Conclusion}

\noindent We have presented several methods to encode sets of finite
vertices as single vertices. In the first part, we used depth-first
spanning trees to encode edges by vertices. As an application we were
able to extend Courcelle's result on the collapse of $\GSO$ to $\MSO$
on sparse hypergraphs from countable hypergraphs to hypergraphs of
arbitrary cardinality.  In the second part we used network flows to
encode arbitrary finite sets by vertices.

Let us mention some open questions.
Considering the first part it would be interesting
to find out whether sparse classes are the only
examples where $\GSO$ collapses to $\MSO$.
\begin{prob}
Is there a class~$C$ that is not $k$-sparse, for any~$k$,
such that over~$C$ every $\GSO$-sentence is equivalent to
an $\MSO$-sentence\??
\end{prob}

The results of the second part are much less complete.
It is unlikely that they are the best possible.
\begin{prob}
Improve Theorem~\ref{Thm: coding sets by vertices}
by allowing
\begin{enuma}
\item more general classes of graphs or hypergraphs\?;
\item more general classes of partial functions $h : \PSet(V) \to V$.
\end{enuma}
\end{prob}
Our results were inspired by work of
Colcombet and L\"oding~\cite{ColcombetLoeding07}.
The question arises of whether we can also generalise
the remaining results of that article.
\begin{prob}
Can we prove Corollary~4.4 of\/~\textup{\cite{ColcombetLoeding07}}
for other graphs than trees\??
\end{prob}

\section*{Acknowledgement}
\noindent I like to thank Bruno Courcelle for his many comments on
earlier versions of this paper.


{\small
\begin{thebibliography}{1}

\bibitem{ColcombetLoeding07}
{\sc T.~Colcombet and C.~L\"oding}, {\em {Transforming Structures by Set
  Interpretations}}, {Logical Methods in Computer Science}, 3 (2007).

\bibitem{Courcelle03}
{\sc B.~Courcelle}, {\em {The monadic second-order logic of graphs
  \smaller{XIV}\kern0.08em: Uniformly sparse graphs and edge set
  quantifications}}, Theoretical Computer Science, 299 (2003), pp.~1--36.

\bibitem{Diestel06}
{\sc R.~Diestel}, {\em {Graph Theory}}, Springer, 3rd~ed., 2006.

\bibitem{GraedelHirschOtto02}
{\sc E.~Gr\"adel, C.~Hirsch, and M.~Otto}, {\em {Back and Forth Between Guarded
  and Modal Logics}}, {ACM Transactions on Computational Logics},  (2002),
  pp.~418--463.

\bibitem{Hodges93}
{\sc W.~Hodges}, {\em {Model Theory}}, Cambridge University Press, 1993.

\bibitem{NesetrilSopenaVignal97}
{\sc J.~Ne{\v s}et{\v r}il, E.~Sopena, and L.~Vignal}, {\em {T-preserving
  homomorphisms of oriented graphs}}, Comment.\ Math.\ Univ.\ Carolinae, 38
  (1997), pp.~125--136.

\bibitem{Seese91}
{\sc D.~Seese}, {\em {The structure of the models of decidable monadic theories
  of graphs}}, Annals of Pure and Applied Logic, 53 (1991), pp.~169--195.

\end{thebibliography}
}
\end{document}